\documentclass{article}
\usepackage{epsfig}
\usepackage{amssymb}
\newcommand{\bfr}{\begin{flushright}}
\newcommand{\efr}{\end{flushright}}
 
\begin{document}
\title{Vacuum polarization around a three-dimensional black hole
%\thanks{Presented at ...}%
}
\author{Kiyoshi Shiraishi\\
%\address{
Akita Junior College, Shimokitade-Sakura, Akita-shi, \\Akita 010,
Japan\\
and\\
Takuya Maki\\
Department of Physics, Tokyo Metropolitan University,\\
Minami-ohsawa, Hachioji-shi, Tokyo 192-03, Japan
%}
}
\date{Class. Quantum Grav. {\bf 11} (1994) 695--699
}
\maketitle
\begin{abstract}
We calculate the Euclidean propagator for a conformally coupled
massless scalar field in the background of the three-dimensional black
hole. The expectation value $\langle\varphi^2\rangle$ in the
Hartle-Hawking state is obtained in the spacetime. 
\end{abstract}
%\PACS{}

Recently, the black hole solution to the three-dimensional Einstein
equations with a negative cosmological constant has been found \cite{1}
and various aspects on the black hole have been examined by many
authors \cite{1,2}. In three-dimensional spacetime, the Einstein
equations in vacua with a negative cosmological constant $-\lambda$ are
reduced to
\begin{equation}
R_{\mu\nu}=-2\lambda g_{\mu\nu}\,. 
\label{eq1}
\end{equation}

The black hole solution to the equations has been obtained \cite{1} in
a simple form. For zero angular momentum, the metric is given by
\begin{equation}
ds^2=-(\lambda r^2-M)dt^2+\frac{dr^2}{\lambda r^2-M}+r^2d\theta^2
\label{eq2}
\end{equation}
where $M$ is the mass of the three dimensional black hole \cite{1,2}.
The black hole horizon is located at $r=r_{H}=(M/\lambda)^{1/2}$.

In this paper, we discuss the vacuum polarization in the non-rotating
black hole background in three dimensions. In order to obtain an
exact expression for $\langle\varphi^2\rangle$ for a conformally coupled
scalar field, we calculate the propagator (two-point function) for the
field in the black hole spacetime with Euclidean signature. The
computation is done by the mode sum method.

We begin with introducing a new coordinate $\rho$ defined by
\begin{equation}
r=r_H\sec \rho\qquad (0\le\rho\le \pi/2)
\label{eq3}
\end{equation}
where $r_H=(M/\lambda)^{1/2}$. The metric takes the following form in
terms of this coordinate:
\begin{equation}
ds^2=\lambda^{-1} (\sec \rho)^2(-M\lambda\sin^2\rho dt^2+d\rho^2+
Md\theta^2)\,.
\label{eq4}
\end{equation}

Substituting the time coordinate $t$ by $-i\tau$, we obtain the
Euclidean metric:
\begin{equation}
ds_E^2=\lambda^{-1}(\sec\rho)^2[d\rho^2+\kappa^2\sin^2\rho d\tau^2+
M d\theta^2] 
\label{eq5}
\end{equation}
where we set $\kappa^2=M\lambda$.

One finds that this Euclidean metric is regular at $\rho=0$, which
corresponds to the horizon $r=r_H$, if the Euclidean time $\tau$ is a
periodic coordinate with period $2\pi/\kappa$. Thus the Hawking
temperature is given by $\kappa/2\pi=(M\lambda)^{1/2}/2\pi$ for the
three-dimensional black hole \cite{1,2}.

Now we consider a scalar field in this spacetime. For a conformally
coupled, massless scalar field in three dimensions, the wave equation
for the scalar field reads
\begin{equation}
\Box\varphi-\frac{1}{8}R\varphi=0 
\label{eq6}
\end{equation}
where the covariant divergence is defined in terms of the background
of the three dimensional black hole. In addition, we must note that the
scalar curvature $R$ takes a constant value, $R=-6\lambda$, in the
spacetime.

The Hartle-Hawking propagator $G_H$ \cite{3} for this scalar field is
the solution to the following equation:
\begin{equation}
\left(\Box-\frac{1}{8}R\right) G_H(x,
x')=-\frac{1}{\sqrt{g}}\delta(x, x')\,. 
\label{eq7}
\end{equation}

Our approach to obtain an explicit form of this propagator is the mode
sum method \cite{4}. The mode functions are the solutions to the
equation (\ref{eq6}) with appropriate boundary conditions. The wave
equation can be solved through separation of variables:
\begin{equation}
\varphi_{mn}(x)=u_{mn}(\rho) e^{im\theta}e^{in\kappa\tau}
\label{eq8}
\end{equation}
where $m$ and $n$ are integers, by which the correct periodicities with
respect to $\theta$ and $\tau$ are satisfied.

Assuming the mode function (\ref{eq8}), we find that the wave equation
(\ref{eq6}) is reduced to a differential equation for the radial
function:
\begin{equation}
\cot\rho\frac{d}{d\rho}\tan\rho\frac{d}{d\rho}u_{mn}(\rho)-
\left[\frac{n^2}{\sin^2\rho}+\frac{m^2}{M}\right]u_{mn}(\rho)=0\,.
\label{eq9}
\end{equation}

We find that the general solution to this equation can be written by
\begin{equation}
u_{mn}(\rho) =(\cos\rho)^{1/2}[\alpha
P^n_{-1/2+im/\sqrt{M}}(\cos\rho)+\beta Q^n_{-1/2+im/\sqrt{M}}(\cos\rho)]
\label{eq10}
\end{equation}
where $P^\mu_\nu(z)$ and $Q^\mu_\nu(z)$ are the Legendre functions and
$\alpha$ and $\beta$ are arbitrary constants.

The propagator $G_H$ is constructed from these mode functions with the
boundary condition that requires the regularity at $\rho=0$ and $\pi/2$.
The normalization is detemined by the Fourier coefficients of the delta
functions and the Wronskian condition with respect to the radial
function. Consequently, the expression for the propagator by mode
summation turns out to be
\begin{eqnarray}
& &G_H(\rho, \tau, \theta;\rho', \tau', \theta')=\frac{\kappa}{2\pi}
\sum_{n=-\infty}^\infty e^{in\kappa(\tau-\tau')}
\sum_{m=-\infty}^\infty
e^{im(\theta-\theta')}(\cos\rho)^{1/2}\nonumber \\ &
&\times(\cos\rho')^{1/2}\frac{(-1)^n}{M}
P^{-n}_{-1/2+im/\sqrt{M}}(\cos\rho_{<})Q^{n}_{-1/2+im/\sqrt{M}}(\cos
\rho_{>})
\label{eq11}
\end{eqnarray}
where $\rho_{<}=\rho$ and $\rho_{>}=\rho'$ if $\rho<\rho'$, while
$\rho_{<}=\rho'$ and $\rho_{>}=\rho$ if $\rho>\rho'$.

This expression can be simplifIed by use of the addition theorem
\cite{5}. We can rewrite the sum over $n$ and then we get
\begin{eqnarray}
& &G_H(\rho, \tau, \theta;\rho', \tau',
\theta')=\frac{1}{4\pi^2r_H}
\sum_{m=-\infty}^\infty
e^{im(\theta-\theta')}(\cos\rho)^{1/2}(\cos\rho')^{1/2}
\nonumber \\ &
&\qquad\times Q_{-1/2+im/\sqrt{M}}(\cos\rho\cos\rho'+\sin\rho\sin\rho'
\cos\kappa(\tau-\tau'))
\label{eq12}
\end{eqnarray}
where $Q_\nu(z)=Q^{\mu=0}_\nu(z)$.

We can further simplify the expression with the help of the integral
representation of the conical function \cite{6}.%
\footnote{There is a misprint in 8.12.4 in their book \cite{6}, the
$\cosh$ in the second line should read $\cos$.}
The sum over $m$ yields delta functions, so the representation becomes
fairly simple. We find
\begin{eqnarray}
& &G_H=\frac{(\cos\rho)^{1/2}(\cos\rho')^{1/2}}{16\sqrt{2}\pi^2r_H}
\sum_{m=-\infty}^\infty\int_{-\infty}^\infty
[\cos m(\theta-\theta')\cos(m\phi/\sqrt{M})d\phi]
\nonumber \\ &
&\times
[\cosh\phi-\cos\rho\cos\rho'-\sin\rho\sin\rho'\cos\kappa(\tau-\tau')
]^{-1/2}\nonumber \\
&
&=\sum_{k=-\infty}^\infty[\sqrt{\lambda}(\cos\rho)^{1/2}(\cos\rho')^{1/2}
][4\sqrt{2}\pi]^{-1}\nonumber \\
& &\times[\cosh\sqrt{M}(\theta-\theta+2\pi
k-\cos\rho\cos\rho'-\sin\rho\sin\rho'\cos\kappa(\tau-\tau'))]^{-1/2}\,.
\label{eq13}
\end{eqnarray}

Now we calculate the vacuum polarization in the black hole spacetime.
The quantum effects around the black hole can be calculated from the
propagator. We first compute the vacuum expectation value
$\langle\varphi^2\rangle$ for a conformally coupled real scalar field,
as the simplest example.

In the Hartle-Hawking vacuum \cite{3}, we take the vacuum polarization
$\langle\varphi^2\rangle$ as the coincidence limit of the
Hartle-Hawking propagator $G_H(x,x)$ obtained above with appropriate
regularization \cite{7,8,9,10,11}. The regularization is done by a
similar method adopted by Frolov et al. \cite{9}.

We find that the propagator takes the form, in the case that the
separation between two points is restricted to the radial direction
(i,e., when $\theta=\theta'$ and $\tau=\tau'$):
\begin{equation}
G_H(\rho, \rho')=\frac{\sqrt{\lambda}(\cos\rho)^{1/2}(\cos\rho')^{
1/2}}{8\pi\sin((\rho-\rho')/2)}+\sum_{k=1}^\infty
\frac{\sqrt{\lambda}(\cos\rho)^{1/2}(\cos\rho')^{
1/2}}{2\sqrt{2}\pi\sqrt{\cosh 2\pi\sqrt{M}k-\cos(\rho-\rho'))}}\,.
\label{eq14}
\end{equation}
Note that the latter summation part in the RHS of (\ref{eq14}) does
not include divergent contribution in the coincidence limit $\rho'
\rightarrow\rho$.

We need the Schwinger-de Witt expansion of the propagator (two-point
function) with respect to the powers of the geodesic distance between
the two points for subtraction of the divergence in the coincidence
limit \cite{4,9}.

The divergent and constant contributions in the Schwinger-De Witt
expansion near the horizon is written by use of the geodesic distance
as \cite{8,9}
\begin{equation}
G^{div}_H(x,x')=\frac{1}{4\pi\sqrt{2\sigma(x,x')}}
\label{eq15}
\end{equation}
where $\sigma=s^2/2$ and $s(x,x')$ is the
geodesic distance between $x$ and $x'$. For the radial scparation, $s$
is given by
\begin{equation}
s(\rho,\rho')=\frac{1}{\sqrt{\lambda}}\int_{\rho'}^\rho
\frac{dy}{\cos
y}=\left[\frac{1}{2\sqrt{\lambda}}\ln\left[\frac{1+\sin y}{1-\sin y}
\right]\right]_{y=\rho'}^{y=\rho}\,.
\label{eq16}
\end{equation}

 Using (\ref{eq14}, \ref{eq15} and \ref{eq16}), we get the final result:
\begin{equation}
\langle\varphi^2\rangle(\rho)=\lim_{\rho'\rightarrow\rho}
(G_H(\rho,\rho')-G_H^{div}(\rho,\rho'))=\sum_{k=1}^\infty
\frac{\sqrt{\lambda}\cos\rho}{4\pi\sinh\pi\sqrt{M}k}
=\langle\varphi^2\rangle(0)\cos\rho\,.
\label{eq17}
\end{equation}

Here the divergent and constant parts of $G_H$ in the first part of the
RHS of (\ref{eq14}) have been exactly cancelled by those of $G^{div}$.
At the edge of the universe, $\rho=\pi/2$, $\langle\varphi^2\rangle$
vanishes as $\cos\rho$. In terms of the original coordinate $r$ in
(\ref{eq2}), it is found that $\langle\varphi^2\rangle$ is proportional
to $1/r$.

We can calculate the propagator and $\langle\varphi^2\rangle$ for a
twisted scalar field which obeys the boundary condition \cite{12}:
\begin{equation}
\varphi(\theta+2\pi)=\varphi(\theta)\,.
\label{eq18}
\end{equation}

The calculation for the twisted field around the three-dimensional
black hole can be done similarly to the previous untwisted case. We
only show the result for the vacuum expectation value
$\langle\varphi^2\rangle_{twisted}$:
\begin{equation}
\langle\varphi^2\rangle_{twisted}(\rho)= \sum_{k=1}^\infty
\frac{\sqrt{\lambda}(-1)^k\cos\rho}{4\pi\sinh\pi\sqrt{M}k}=
\langle\varphi^2\rangle_{twisted}(0)\cos\rho\,.
\label{eq19}
\end{equation}

%**************************************************************
\begin{figure}[ht]
\begin{center}
\includegraphics[width=6cm]{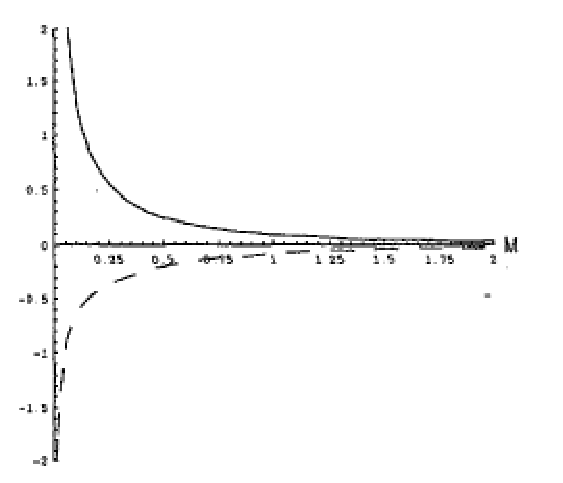}
\caption{The magnitude of the vacuum polarization on the horizon as a
function of $M$. The solid curve represents for
$4\pi\langle\varphi^2\rangle(0)/\sqrt{\lambda}$, while the dashed
curve for $4\pi\langle\varphi^2\rangle_{twisted}(0)/\sqrt{\lambda}$.}
\label{f1}\end{center}
\end{figure}
%**************************************************************

The numerical results im shown in Figure~1. For large $M$, the absolute
value of $\langle\varphi^2\rangle$ on the horizon is dumped as
$\exp(-\pi\sqrt{M})$, while in the limit of $M\rightarrow 0$,
$\langle\varphi^2\rangle$ diverges.

In this paper, we have calculated the Hartle-Hawking propagator for a
conformally coupled massless scalar field in three-dimensional black
hole spacetime with Euclidean signature. Using the exact form of the
propagator, we have obtained the vacuum value $\langle\varphi^2\rangle$
for untwisted and twisted scalar fields in the Hartle-Hawking vacuum.
Its dependence on the radial coodinate has been found as
$\langle\varphi^2\rangle\approx\cos\rho\approx 1/r$. The mass
dependence of $\langle\varphi^2\rangle(0)$ has
been numerically evaluated and shown in Figure~1.

In the limit of small mass, the amount of the vacuum fluctuation becomes
unlimitedly large, according to our result. We must consider the back
reaction to the metric in such a case. The effect of the vacuum
fluctuation may have much importance on the final stage of the black
hole evaporation.

We discussed only the conformally coupled massless scalar field. One
may wish to extend the analyses in the present paper to the general
couplings and masses. The rotation as well as the charge of the black
hole in three dimensions will change the behaviour of the quantum
fields. These are interesting subjects worth studying. The
thermodynamics of the three-dimensional black holes and the effect of
the back reaction due to the quantum effects should also be investigated
in the future.

%%%%%%%%%%%%%%%%%%%%%%%%%%%%%%%%%%%%%%%%%%%%%% 

%%%%%%%%%%%%%%%%%%%%%%%%%%%%%%%%%%%%%%%%%%%%%
\end{document}